\documentclass[journal]{IEEEtran}
\ifCLASSINFOpdf
\else
\fi
\usepackage{color}
\usepackage{graphicx}
\usepackage{booktabs}
\usepackage{amsmath}
\usepackage{alltt}
\usepackage{cite}
\begin{document}
%
\title{Purcell Effect in the Stimulated and Spontaneous Emission Rates of Nanoscale Semiconductor Lasers}
%
%
%
%
\author{Bruno~Romeira
        and~Andrea~Fiore
\thanks{B. Romeira and A. Fiore are with the Department of Applied Physics and Institute for Photonic Integration, Eindhoven University of Technology, Postbus 513,
5600 MB, Eindhoven, The Netherlands (e-mail: bruno.romeira@inl.int; a.fiore@tue.nl).}
\thanks{B. Romeira current address: Department of Nanophotonics, International Iberian Nanotechnology Laboratory, Av. Mestre Jos\'{e} Veiga, Braga 4715-330, Portugal.}
\thanks{This work was supported by NanoNextNL, a micro- and nanotechnology program of the Dutch Ministry of Economic Affairs and Agriculture and Innovation and 130 partners and by the NWO Zwaartekracht program "Research Center for Integrated Nanophotonics". B.R. acknowledges the financial support of the Marie Sk{\l}odowska Curie IF fellowship NANOLASER (2014-IF-659012).}}

\maketitle

\begin{abstract}
Nanoscale semiconductor lasers have been developed recently using either metal, metallo-dielectric or photonic crystal nanocavities. While the technology of nanolasers is steadily being deployed, their expected performance for on-chip optical interconnects is still largely unknown due to a limited understanding of some of their key features. Specifically, as the cavity size is reduced with respect to the emission wavelength, the stimulated and the spontaneous emission rates are modified, which is known as the Purcell effect in the context of cavity quantum electrodynamics. This effect is expected to have a major impact in the 'threshold-less' behavior of nanolasers and in their modulation speed, but its role is poorly understood in practical laser structures, characterized by significant homogeneous and inhomogeneous broadening and by a complex spatial distribution of the active material and cavity field. In this work, we investigate the role of Purcell effect in the stimulated and spontaneous emission rates of semiconductor lasers taking into account the carriers' spatial distribution in the volume of the active region over a wide range of cavity dimensions and emitter/cavity linewidths, enabling the detailed modeling of the static and dynamic characteristics of either micro- or nano-scale lasers using single-mode rate-equations analysis. The ultimate limits of scaling down these nanoscale light sources in terms of Purcell enhancement and modulation speed are also discussed showing that the ultrafast modulation properties predicted in nanolasers are a direct consequence of the enhancement of the stimulated emission rate via reduction of the mode volume.
\end{abstract}


%
\IEEEpeerreviewmaketitle

\section{Introduction}\label{sec_introduction}
\IEEEPARstart{N}{anolasers}, with dimensions smaller than the emitted wavelength, show great potential due to their unique features including ultra-small footprint, high-speed modulation and unprecedented low energy budgets. This can have a crucial impact, not only in future optical interconnects and communications systems working at ultralow energy per bit levels ($<$10 fJ/bit \cite{Miller2010}) and at tens of gigabit per second (Gb/s) speeds, but also in sensing applications \cite{Kita2010,Choi2016}. Additionally, a great variety of physical quantum phenomena including photon bunching and superradiant emission \cite{Jahnke2016} can be experimentally studied in detail taking advantage of the developed nanolasers.

Several research groups recently succeeded in achieving lasing in a wide range number of photonic crystal \cite{Colombelli2003,Park2004,Altug2006,Ellis2011,Takeda2013}, metallo-dielectric \cite{Hill2007,Nezhad2010,Lee2011,Ding2012,Ding2013a} and plasmonic \cite{Oulton2009,Ning2010,Lu2012,Sidiropoulos2014,Bermudez-Urena2017} nanocavities. It is important to note that although ultra high-speed operation ($>0.1$ THz) has been predicted in nanocavity lasers, their respective  modulation properties have only been experimentally reported in very few cases, including the work of Altug \emph{et al.} that demonstrated direct modulation speeds exceeding 100 GHz in an optically pumped photonic crystal nanocavity laser \cite{Altug2006}, and the work of Sidiropoulos \emph{et al.} which reported pulses shorter than 800 fs from optically pumped hybrid plasmonic zinc oxide (ZnO) nanowire lasers \cite{Sidiropoulos2014}. While the technology of nanolasers is steadily being deployed (e.g., see recent reviews in \cite{Hill2014,Gu2014,Gu}), such high-speeds have not been experimentally tested in electrically driven nanolasers and their expected performance for on-chip and intra-chip optical interconnects is still largely unknown due to a limited understanding of some of their key features, specifically the modulation dynamic properties. Besides the many technological challenges \cite{Ding2013a}, as the cavity size of a nanolaser becomes of the order of the emission wavelength, new physical phenomena such as the Purcell effect \cite{Purcell1946} play a major role in some of the expected unique properties of nanolasers, including lasing at extremely low threshold values \cite{Baba2003}\cite{Ellis2011,Wu2015}, the possibility of realizing 'threshold-less' lasers \cite{Baba1991,Khajavikhan2011,Rieto2015} and ultrafast modulation speeds \cite{Lau2009,Suhr2010,Ni2012,Altug2006,Sidiropoulos2014,Ding2015}.

E. M. Purcell described in 1946 that for a system coupled to an electromagnetic resonator the spontaneous emission probability is increased over its bulk value, and the recombination time reduced, by a factor \cite{Purcell1946}:

\begin{equation}\label{eq_purcell_factor}
  F_{P} = \frac{3\lambda_c^{3}}{4\pi^{2}}\frac{Q}{V}
\end{equation}
which is now called the Purcell factor. In Eq. (\ref{eq_purcell_factor}) the parameter $V$ is the volume of the resonant mode, $Q$ its quality factor, and $\lambda_c$ the wavelength in the material ($\lambda_c=\lambda_0/n_{ra}$, where $n_{ra}$ is the refractive index of the medium). We note that Eq. (\ref{eq_purcell_factor}) considers a cavity whose fundamental mode is resonant with the transition frequency, for a dipole aligned with the polarization of this cavity mode and located at position of maximum field, and for an emitter linewidth that is narrow compared with the cavity linewidth. This makes the experimental observation of this effect relatively challenging, specifically in the field of optics where a significant increase in the emission rate requires optical resonators that are able to confine light down to dimensions comparable to the wavelength and store it for a long time.

In a seminal paper in 1998 \cite{Gerard1998}, J. M. G\'{e}rard \emph{et al.} demonstrated the Purcell enhancement of the spontaneous emission by semiconductor quantum dots in a monolithic optical microcavity paving the way to fascinating quantum electrodynamics experiments on single solid-state quantum emitters in microcavities \cite{Solomon2001}. Since then, the processing technology has matured sufficiently to enable the fabrication of nanocavities with high quality-factors and low mode volumes and demonstrate the Purcell enhancement in a variety of dielectric or metallic cavities \cite{Gerard1999,Englund2007,Halas2011,Khurgin2015}.

In semiconductor micro- and nano-lasers, a few seminal theoretical works in the early 90's already predicted \cite{Yokoyama1989,Yamamoto1991,Bjork1991,Yamamoto1992,Yokoyama1992}, using rate-equation analysis, that microcavity lasers taking advantage of an enhancement of the emission rate could display unique and novel properties, including a low threshold current, the disappearance of the lasing threshold in the input-output curve and the absence of relaxation oscillations. After the development of the first micro- and nano-cavity lasers showing some of these unique properties \cite{Baba1991,Altug2006,Khajavikhan2011,Rieto2015}, a wide range of theoretical models have been proposed to study the corresponding static and dynamical characteristics. These works include the analysis of the modulation speed in nanocavity light emitting (LED) devices and nanolasers \cite{Lau2009,Suhr2010,Ni2012} as a function of the mode volume, $V$, spontaneous emission factor, $\beta$, and Purcell factor, $ F_{P}$. More detailed rate-equation models for plasmonic nanolasers have been proposed that include description of metal effects and take into account the inhomogeneity and dispersion of the cavity media \cite{Chang2008,Chang2009}. Lastly, in the case that the number of photons in the nanolaser cavity is very low, a quantum description of the nanolaser may be required, as proposed by several authors \cite{Gies2007,Moelbjerg2013,Lorke2013,Chow2014}.

It is noteworthy that, contrary to the seminal work of Yokohama \emph{et al.} \cite{Yokoyama1989}, most of recently reported rate-equation models mainly focus on the enhancement of spontaneous emission \cite{Ding2012,Lau2009,Suhr2010}, neglecting that stimulated emission is directly linked to spontaneous emission, as it is readily seen in the Einstein's relations or by the derivation of the light-matter interaction in the quantized field picture. Several experimental studies also report the enhancement of the stimulated emission, like for example in microdroplets \cite{Campillo1991}, microlasers \cite{Djiango2008}, and nanowire lasers \cite{Wei2015}, confirming that it should be treated on the same footing as the spontaneous emission. In the case of a nanolaser, this can result in a Purcell enhancement of the stimulated emission which influences the threshold of the laser, as theoretically investigated in the case of spectrally-narrow emitters \cite{Gregersen2012}, and as reported in a recent experimental work on subwavelength red-emitting hybrid plasmonic lasers \cite{Liu2016}. Another recurring assumption in nanolaser models is that the Purcell factor, $F_P$, and in some cases the spontaneous emission coupling factor, $\beta$, can be treated as adjustable independent phenomenological parameters. Although this approach can provide a reasonable qualitative description of a nanolaser, in an experimental device only the cavity dimensions and emitter/cavity linewidths can be controlled. Therefore, $\beta$ and $F_p$ are not device adjustable parameters but come as a result of the emission processes occurring in a nanolaser. It is also clear that the Purcell factor and the $\beta$ factor are in general independent quantities. The Purcell factor presented in Eq. (\ref{eq_purcell_factor}) can be more generally defined as:

 \begin{equation}\label{eq_purcell_bulk}
  F = \frac{R_{sp,cav}}{R_{bulk}}
\end{equation}
where $R_{sp,cav}$ is the spontaneous emission rate into the cavity mode per unit time and $R_{bulk}$ is the total spontaneous emission rate per unit time in the bulk medium in the absence of a cavity. The Purcell factor, $F$, in Eq. (\ref{eq_purcell_bulk}), becomes equal to $F_P$, Eq. (\ref{eq_purcell_factor}), in the case of an ideally matched emitter. The spontaneous emission coupling factor, $\beta$, is defined as the fraction of spontaneously emitted photons which are coupled to the cavity mode, and can be written as:

 \begin{equation}\label{eq_beta}
  \beta = \frac{R_{sp,cav}}{R_{sp,cav}+R_{l}}
\end{equation}
where $R_{l}$ is the rate of emission per unit time into other modes. As can be immediately seen from  Eq. (\ref{eq_purcell_bulk}) and Eq. (\ref{eq_beta}), the $\beta$ value can be substantially increased via the suppression of $R_{l}$, even when $R_{sp,cav}=R_{bulk}$, i.e. in the absence of Purcell enhancement, as discussed for example in \cite{Gerard2003}.

Finally, since the active medium in nanolasers mostly consists of bulk or multi-quantum well (MQW) semiconductors, inhomogeneous and homogeneous broadenings should be taken into account, particularly when the nanolasers operate at room temperature. In the situation of a gain medium described by a broad emitter, as outlined for example in \cite{Auffeves2009,Gu2013}, the linewidth broadening typically overcomes the effect of a much narrower cavity linewidth, and consequently the cavity $Q$ has negligible effect on the spontaneous emission rate, a case not described by Eq. (\ref{eq_purcell_factor}). Therefore, in nanocavity lasers this may result in much lower overall spontaneous emission rates than that predicted by Eq. (\ref{eq_purcell_factor}) for a narrow emitter. Additionally, as discussed recently in the case of a metal-dielectric nanopillar cavity \cite{Dolores-Calzadilla2017}, the carrier distribution can be non-uniform over the mode volume, which can further reduce the spontaneous emission rate.

In this work, we take all these effects into account and present a single-mode rate-equation model that considers the Purcell enhancement of both spontaneous and stimulated emission rates on the same footing. Using this model, we investigate in detail the static and dynamic characteristics of electrically-pumped metallo-dielectric cavity nanolasers, including threshold current and modulation speed properties. The treatment presented here is fundamentally different from the rate-equation analysis reported before due to the following combined key aspects:

\begin{description}
  \item[i)] Only the physical properties of the nanolasers, specifically the gain material and cavity, are used to fully describe their static and dynamic characteristics, avoiding the \emph{ad-hoc} introduction of the spontaneous emission factor, $\beta$, or the Purcell factor, $F_P$;
  \item[ii)] Spontaneous and stimulated emission rates are treated on the same footing which leads to a Purcell enhancement of both radiative processes;
  \item[iii)] The model can describe either macro-, micro- or nano-lasers over a wide range of cavity dimensions and emitter/cavity relative linewidths (including, but not limited to, photonic crystal/metallic cavities and quantum dot/well/bulk gain materials);
  \item[iv)] The model accounts for the spatial and spectral overlap between carriers and photons.
\end{description}

The paper is organized as follows. In section \ref{sec_purcell}, we write the stimulated and spontaneous emission rates for a homogeneously broadened two-level atom in a resonant cavity using the Fermi's golden rule. We consider the Purcell enhancement in two general situations: 1) spectrally-narrow emitter (much narrower than the single cavity mode) and 2) broad emitter. In section \ref{sec_rate_equations}, we introduce the detailed single-mode rate-equation model and extend our treatment to account for the inhomogeneous broadening of the carriers, and the carriers' spatial distribution in the volume of the active region in the case of nanocavity lasers. In sections \ref{sec_static} and \ref{sec_dynamic}, we analyze the static and dynamic characteristics, respectively, of both electrically pumped micropillar lasers and nanopillar metal-cavity lasers, considering the most common situation of lasers operating at room-temperature and employing a bulk gain active medium (e.g. InGaAs), i. e., with a gain spectrum much broader than the cavity mode. The expected performance in terms of threshold current and high-speed modulation is discussed in detail. The ultimate limits of scaling down these nanoscale light sources in terms of Purcell enhancement of the emission are also discussed.

\section{Purcell enhancement of the spontaneous and stimulated emission rates}\label{sec_purcell}

\subsection{Stimulated and spontaneous emission from Fermi's Golden rule}

The rate of photon emission for a homogeneously broadened two-level atom in a resonant cavity is derived directly from Fermi's golden rule \cite{Ujihara1993}:

\begin{equation}\label{eq_fermi_golden_rule}
  R_{em}=\frac{2\pi}{\hbar^{2}}\int_{0}^{\infty} | \langle f |H| i \rangle|^{2}\rho(\omega)L(\omega) d\omega
\end{equation}
where $\rho(\omega)$ is the density of optical states per unit of angular frequency $\omega$, $L(\omega)$ the homogeneous broadening lineshape, $H$ the atom-field interaction hamiltonian, and $i, f$ the initial and final states of the transition. The lineshapes for the cavity and the emitter are both typically given by Lorentzians. A detailed derivation of Eq. (\ref{eq_fermi_golden_rule}), based on the density matrix approach and an artificial discretization of the density of optical states can be found in \cite{Ujihara1993}, and is employed in \cite{Yokoyama1995} in the investigation of the spontaneous emission in optical microcavities.

Considering an electric dipole transition and a single cavity mode, the matrix element is given by:
\begin{equation}\label{eq_matrix_element}
  | \langle f |H| i \rangle|=E_{0}(\vec{r}_{em})\hat{e}\cdot\vec{d}_{if}\sqrt{N_{ph}+1}
\end{equation}
where $N_{ph}$ is the number of photons in the mode, $E_{0}(\vec{r}_{em})=\sqrt{(\hbar\omega/2\varepsilon_{0}\varepsilon_{ra}V)}e(\vec{r}_{em})$ is the magnitude of the field per photon at the position of the emitter $\vec{r}_{em}$, $\hat{e}$ is a unit vector indicating its polarization, $\varepsilon_{0}$ is the dielectric permitivity of free space, $\varepsilon_{ra}$ is the relative dielectric constant in the active material and $V$ the cavity mode volume. In Eq. (\ref{eq_matrix_element}), we note that the term proportional to $N_{ph}$ denotes stimulated emission, while the term proportional to 1 represents spontaneous emission. The adimensional mode function $e(\vec{r})$ is normalized to be $|e(\vec{r})|_{max}=1$, so that for a point-like, optimally positioned emitter, we obtain:

 \begin{equation}\label{eq_matrix_element_ideal_case}
  \langle f |H| i \rangle|=\sqrt{\left( \frac{\hbar\omega}{2\varepsilon_{0}\varepsilon_{ra}V} \right)}\hat{e}\cdot\vec{d}_{if}\sqrt{N_{ph}+1}
\end{equation}
which is in agreement with the usual notation \cite{Gerry2004}. Finally, the atomic dipole moment is defined as:

\begin{equation}\label{eq_dipole_moment}
  \vec{d}_{if}=|\langle\Psi_{i}|\vec{d}|\Psi_{f}\rangle|
\end{equation}
where $\vec{d}$ is the dipole operator, and $\Psi_{i(f)}$ is the upper (lower) level wavefunction of the atom.

The mode volume, $V$, is defined by the energy normalization condition of the field per photon, $E_0$ , which in the case of a dielectric, non-dispersive cavity reads:
\begin{equation}\label{eq_energy_normalization}
  \int 2\varepsilon(\vec{r})|E_{0}(\vec{r})|^{2}d^{3}\vec{r}=\hbar\omega \\
\end{equation}
The mode volume is therefore given by:
\begin{equation}\label{eq_mode_volume}
  V = \int\frac{\varepsilon_{r}(\vec{r})}{\varepsilon_{ra}}|e(\vec{r})|^{2}d^{3}\vec{r}
\end{equation}
In the case where the dielectric constant is uniform in the cavity, this simplifies to $ V = \int|e(\vec{r})|^{2}d^{3}\vec{r}$ and the mode volume is close to the physical cavity volume. Note that the normalization condition in Eq. (\ref{eq_energy_normalization}) changes in the situation of a cavity with metal boundaries, since the energy in the field and the kinetic energy of the electrons both have to be properly accounted for, as thoroughly discussed in \cite{Khurgin2014}. This changes the value of $V$ but does not qualitatively modify the discussion below.

\begin{figure}
  \centering
  \includegraphics[width=3.0in]{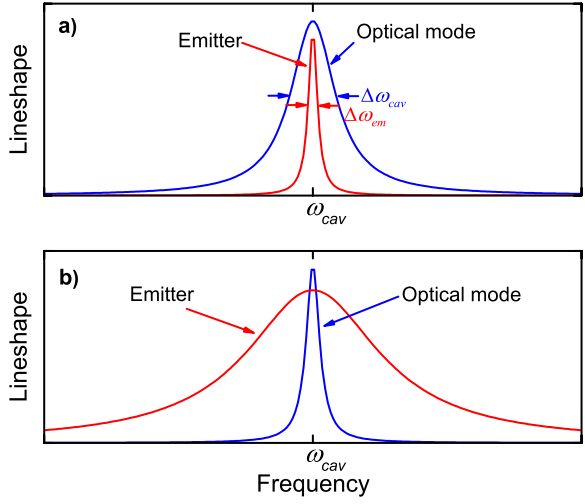}\\
  \caption{Schematic representation of two situations showing the relation between the cavity resonance and the emitter's transition spectrum. (a) The single cavity mode resonance width is broader than the emitter's transition width. (b) The emitter's width is broader than the single cavity mode width.}\label{fig_cavity_emitter}
\end{figure}

In order to relate to the literature on Purcell-enhanced spontaneous emission, we now apply Eq. (\ref{eq_fermi_golden_rule}) and the matrix element for a point-like, optimally positioned emitter, Eq. (\ref{eq_matrix_element_ideal_case}), and derive the stimulated and spontaneous emission rates in the cavity mode considering two situations. The first case analyzed, depicted in Fig \ref{fig_cavity_emitter}(a), considers that the emitter has a delta-function-like spectral width in comparison to the cavity. This is the standard example in which large Purcell factors \cite{Purcell1946} are observed and applies for example to a quantum dot (QD) emitter at cryogenic temperatures where the homogeneous linewidth can be made smaller than 0.1 meV (for a review see \cite{Pelton2015}), a case treated in many quantum optics textbooks \cite{A.M.Fox2006}. In a second case, schematically represented in Fig \ref{fig_cavity_emitter}(b), we analyze the situation when the cavity linewidth is a delta function as compared to the emitter linewidth, which is the standard situation of the majority of micro- and nanoscale semiconductor lasers employing a bulk (or MQW) type of emitter operating at room temperature. In a third case, not explicitly treated here, in which the emitter and cavity linewidths are comparable, Eq. (\ref{eq_fermi_golden_rule}) has to be integrated over both lineshapes and the matrix element. This is discussed, for example, in \cite{Meldrum:10} for the case of an ensemble of finite-linewidth quantum dot emitters in a microcavity where the cavity and QDs both have a Lorentzian profile.

\subsection{Emitter narrow, cavity broad $(\Delta\omega_{em}\ll\Delta\omega_{cav})$}

Using Eqs. (\ref{eq_fermi_golden_rule})-(\ref{eq_matrix_element_ideal_case}), we analyze the spontaneous and stimulated emission processes in the case where $\rho(\omega)$ is nonzero only in a narrow frequency region (single optical mode in the gain spectrum), so that the matrix element is independent of frequency and can be taken out of the integral in Eq. (\ref{eq_fermi_golden_rule}). An average dipole moment, $d_{if}=\langle\hat{e}\cdot\vec{d}_{if}\rangle$, is also assumed in the following, for the cases where the emitters have dipole moments oriented along different directions. Thus, the photon creation rate by spontaneous and stimulated emission for a single atom in the cavity mode for a point-like, optimally positioned emitter (still assuming $|e(\vec{r}_{em})|=1$) becomes:

\begin{equation}\label{eq_rsp_cav_stimulated}
  R_{sp,st,cav}= \frac{\pi}{\hbar\varepsilon_{0}\varepsilon_{ra}}\frac{d_{if}^{2}}{V}(N_{ph}+1)\int_{0}^{\infty} \rho(\omega)L(\omega) d\omega
\end{equation}
In the case of a narrow emitter, that is, the linewidth of the emitter is much smaller than the linewidth cavity, $\Delta\omega_{em}\ll\Delta\omega_{cav}$ (Fig. \ref{fig_cavity_emitter}(a)), and assuming the emitter is resonant with the cavity mode $(\omega=\omega_{cav})$ the density of optical modes (assumed Lorentzian) is given by $\rho(\omega_{em})\simeq 2/\pi\Delta\omega_{cav}$ at the emitter's frequency. The stimulated and spontaneous emission rate in the cavity mode (approximating $L(\omega)$ by a Dirac delta function) becomes:

\begin{equation}\label{eq_rsp_cav_stimulated_narrow_emitter}
  R_{sp,st,cav} \simeq \frac{2}{\hbar\varepsilon_{0}\varepsilon_{ra}}d_{if}^{2}\frac{Q}{V}(N_{ph}+1)
\end{equation}
where $Q=\omega_{cav}/\Delta\omega_{cav}$ is the quality factor of the cavity mode, $\omega_{cav}$ is the cavity frequency. Dividing Eq. (\ref{eq_rsp_cav_stimulated_narrow_emitter}) by the emission in the bulk, $R_{bulk}$, for the case of spontaneous emission ($N_{ph}=0$), we obviously retrieve the Purcell factor of Eq. (\ref{eq_purcell_factor}) \cite{Purcell1946}. We note that the well-known $Q/V$ factor in Eq. (\ref{eq_purcell_factor}) and Eq. (\ref{eq_rsp_cav_stimulated_narrow_emitter}) is only applicable to this case of narrow emitter, which is a very specific situation in nanolaser devices. As an example, this would correspond to a single quantum-dot operating at cryogenic temperature. In this case, the photon creation rate by spontaneous and stimulated emission both depend directly from the $Q/V$ ratio, meaning that the stimulated emission can be also Purcell enhanced, a case studied by Gregersen \emph{et al.} \cite{Gregersen2012}.

\subsection{Emitter broad, cavity narrow $(\Delta\omega_{em}\gg\Delta\omega_{cav})$}\label{rsp_st_broad_emitter}

In the most realistic case, the emitter width is broader than the single cavity mode width. This occurs in the thermally broadened gain curve for bulk or quantum well active layers and in quantum dots where a large homogeneous broadening occurs at high temperature or under electrical pumping conditions \cite{Borri2001}. In these cases, $\Delta\omega_{em}\gg\Delta\omega_{cav}$, Fig. \ref{fig_cavity_emitter}(b), using the same procedure as in the previous sub-section, assuming that the homogeneous broadening, $L(\omega)$, is a Lorentzian centered at $\omega_{cav}$ (i.e., $L(\omega_{cav})= 2/\pi\Delta\omega_{em}$) and approximating $\rho(\omega)$ by a Dirac delta function centered at $\omega=\omega_{cav}$, the stimulated and spontaneous emission rate in the cavity mode now becomes:

\begin{equation}\label{eq_rsp_cav_stimulated_broad_emitter}
  R_{sp,st,cav} \simeq \frac{2}{\hbar\varepsilon_{0}\varepsilon_{ra}}d_{if}^{2}\frac{\omega_{cav}}{\Delta\omega_{em}}\frac{1}{V}(N_{ph}+1)
\end{equation}
We note that in this case, both stimulated and spontaneous emission processes depend on the emitter's linewidth, $\Delta\omega_{em}$, and still depend inversely on $V$ but not on the cavity's $Q-$factor (the same expression as Eq. (\ref{eq_rsp_cav_stimulated_broad_emitter}) was previously derived with a Master equation approach in the case of spontaneous emission \cite{Auffeves2009}). The $1/V$ dependence comes directly from the dependence of the emission rate on the field per photon.

\section{Rate-equation analysis}\label{sec_rate_equations}

In this section, we introduce the detailed single-mode rate-equation model considering the practical situation of a semiconductor laser employing a gain medium with homogeneous broadening larger than the cavity linewidth, the case described in Eq. (\ref{eq_rsp_cav_stimulated_broad_emitter}) for a single emitter. Firstly, we derive the rate equations for the case of a semiconductor active medium with broad homogeneous broadening consisting of an atomic ensemble of identical incoherent (wide) solid-state emitters at room temperature. Secondly, we consider the inhomogeneous broadening of the electronic states. In both cases, the typical situation of macro- and microscale lasers (e.g. Fabry-P\'{e}rot, distributed feedback or vertical-cavity surface-emitting cavity lasers) is assumed, where the mode field can be considered uniform in the active region. As a third step, we derive the rate equations for the case of nanoscale (wavelength and sub-wavelength scale) semiconductor lasers in which the field can vary substantially over the volume of the active region.

\subsection{Rate equations with homogeneous broadening}

Starting from Eq. (\ref{eq_rsp_cav_stimulated_broad_emitter}) that describes the photon creation rate by spontaneous and stimulated emission for a single atom, we now consider the realistic situation of a laser active material consisting of an atomic ensemble of incoherent (wide) solid-state emitters at room-temperature in which collective radiant effects (e.g. superradiance \cite{Jahnke2016}) can be considered negligible as the decoherence time of atoms is much shorter than all relevant dynamics. We also initially assume that all atoms have the same transition frequency and are positioned at the antinode of the cavity field so that $|e(\vec{r}_{em})|=1$. The total rate of increase in photon number is then given by $R_{sp,st,cav}^{tot}=R_{sp,st,cav}N_2$, where $N_{2}$ is the number of atoms in the upper state. Since the absorption rate per atom is written similarly to the stimulated emission rate, the net rate of variation in photon number is therefore given by the rate equation:

\begin{multline}\label{eq_rate_equation_total_rate_broad_emitter}
 \frac{dN_{ph}}{dt} = \frac{2}{\hbar\varepsilon_{0}\varepsilon_{ra}}d_{if}^{2}\frac{\omega_{cav}}{\Delta\omega_{em}}\frac{1}{V}(N_{2}-N_{1})N_{ph} \\+\frac{2}{\hbar\varepsilon_{0}\varepsilon_{ra}}d_{if}^{2}\frac{\omega_{cav}}{\Delta\omega_{em}}\frac{1}{V}N_{2}
\end{multline}
where $N_1$ is the number of atoms in the lower state. In order to write a rate equation for the photon, $n_{ph}$ and carrier, $n$, densities we define, $n_{ph}=N_{ph}/V$, and $n_{i}=N_{i}/V_a$ (where $V_a$ is the volume of the active material), respectively. We then arrive at the usual single-mode rate equation for the photon density describing net gain and spontaneous emission:

\begin{multline}\label{eq_rate_equation_photon_simple}
 \frac{dn_{ph}}{dt} = \frac{2}{\hbar\varepsilon_{0}\varepsilon_{ra}} d_{if}^{2}\frac{ \omega_{cav}}{\Delta\omega_{em}}\frac{V_a}{V}(n_{2}-n_{1})n_{ph} \\+\frac{2}{\hbar\varepsilon_{0}\varepsilon_{ra}}d_{if}^{2}\frac{\omega_{cav}}{\Delta\omega_{em}}\frac{V_a}{V^2} n_{2}
\end{multline}
where $V_a/V$ can be defined as the confinement factor, $\Gamma$. In the right side of Eq. (\ref{eq_rate_equation_photon_simple}) we recognize the usual modal gain and spontaneous emission terms typically appearing in single mode rate equations. Both terms depend on the mode volume, which means that a reduction $V$ results not only in an increase of the spontaneous emission into the cavity mode, but also in an increase of the modal gain. Finally, we note that the spontaneous emission term has a $1/V^{2}$ dependence. One $1/V$ term comes from the dependence on the field per photon as already discussed and presented in Eq. (\ref{eq_rsp_cav_stimulated_broad_emitter}). The second $1/V$ dependence is just a result of the use of volume densities in the rate equation.

\subsection{Rate equations with inhomogeneous and homogeneous broadening}

In Eq. (\ref{eq_rate_equation_photon_simple}), only homogeneous broadening has been considered. However, in the case of a semiconductor active medium where emitters are spectrally dispersed, the inhomogeneous broadening of the electronic states must be described by integrating Eq. (\ref{eq_fermi_golden_rule}) over the band structure. While the case of a bulk medium is explicitly treated here, a similar derivation can be done for quantum well, quantum wire or quantum dot active regions. We assume for simplicity a single valence band and a parabolic dispersion of the conduction and valence bands, characterized by temperature-independent effective masses, $m_{e}$ and $m_{h}$, respectively, and neglect the dependence of the matrix element on the wave vector $k$. Following the usual procedure in semiconductor laser theory, we can introduce the joint density of states per unit frequency and volume:

\begin{equation}\label{eq_density_states}
  \rho_{j}(\omega_{vc})= \frac{1}{2\pi^{2}}\left( \frac{2m_{r}}{\hbar}\right)^{3/2}\sqrt{\omega_{vc}-\omega_{g}}
\end{equation}
where $\omega_{vc}$ is the frequency corresponding to a vertical electron-hole transition, $m_{r}=\frac{m_{e}m_{h}}{m_{e}+m_{h}}$, is the reduced mass, and $\omega_{g}$ is the frequency corresponding to the bandgap. In the limit $\Delta\omega_{em}\gg\Delta\omega_{cav}$, assuming that electron and hole populations are in quasi-equilibrium so that they can be described by Fermi-Dirac distributions, and considering the medium as dispersionless, the stimulated and spontaneous emission rates into the cavity mode per unit time and volume, $r_{sp,st,cav}\equiv\frac{R_{sp,st,cav}}{V_a}$, obtained by integrating Eq. (\ref{eq_fermi_golden_rule}) over the band structure, become:

\begin{multline}\label{eq_st_sp_emission_rate_density}
  r_{sp,st,cav}= \frac{\pi}{\hbar\varepsilon_{0}\varepsilon_{ra}}d_{if}^{2}\frac{\omega_{cav}}{V}(N_{ph}+1) \\ \times\int_{\omega_{g}}^{\infty} \rho_{j}(\omega_{vc}) L(\omega_{cav}-\omega_{vc}) f_{c}(1-f_{v}) d\omega_{vc}
\end{multline}
where $f_{c}$, $f_{v}$ are the Fermi distribution functions of electrons calculated at the conduction and valence energies, respectively.

Finally, in order to find the net stimulated emission, $r_{net}=r_{st}-r_{abs}$, we need to consider the expression for the absorption rate, which is similar to the one obtained for the stimulated emission into the cavity mode, Eq. (\ref{eq_st_sp_emission_rate_density}), but with the exchange of the occupation probabilities. Therefore, the net stimulated emission per unit time and volume, $r_{net}$, reads:

\begin{multline}\label{eq_net_gain_density}
  r_{net}\equiv\gamma_{net}n_{ph}= \frac{\pi}{\hbar\varepsilon_{0}\varepsilon_{ra}}d_{if}^{2}\omega_{cav}n_{ph} \\ \times\int_{\omega_{g}}^{\infty} \rho_{j}(\omega_{vc}) L(\omega_{cav}-\omega_{vc}) (f_{c}-f_{v}) d\omega_{vc}
\end{multline}
We note that this equation directly provides the textbook expression for the material gain in the case of a macroscopic cavity \cite{Coldren2012}. The spontaneous emission rate into the cavity mode per unit time and volume is taken from Eq. (\ref{eq_st_sp_emission_rate_density}) by simply setting $N_{ph}=0$:

\begin{multline}\label{eq_sp_density}
  r_{sp,cav} \equiv \frac{\gamma_{sp,cav}}{V} = \frac{\pi}{\hbar\varepsilon_{0}\varepsilon_{ra}}d_{if}^{2}\frac{\omega_{cav}}{V} \\ \times\int_{\omega_{g}}^{\infty} \rho_{j}(\omega_{vc}) L(\omega_{cav}-\omega_{vc}) f_{c}(1-f_{v}) d\omega_{vc}
\end{multline}

In  Eqs. (\ref{eq_net_gain_density}) and (\ref{eq_sp_density}) we introduced $\gamma_{net}$ and $\gamma_{sp,cav}$, respectively, which are volume independent to explicitly show the mode volume dependence on the rates. Lastly, we can write the single-mode rate equations for carrier density, $n$, and photon density, $n_{ph}$, assuming an electrically pumped semiconductor laser:

\begin{eqnarray}
  \frac{dn}{dt} &=& \frac{\eta_{i} I}{q V_a}-r_{nr}-r_{l}-\frac{\gamma_{sp,cav}}{V}-\gamma_{net}n_{ph} \label{eq_rate_equations_1}\\
  \frac{dn_{ph}}{dt} &=& \frac{V_a}{V} \gamma_{net}n_{ph}+\frac{V_a}{V} \frac{\gamma_{sp,cav}}{V}-\frac{n_{ph}}{\tau_{p}}\label{eq_rate_equations_2}
\end{eqnarray}
where we have introduced the recombination rates per unit time and volume, $r_{nr}$ and $r_l$, describing the non-radiative recombination rate and the radiative recombination rate into the leaky modes, respectively. The non-radiative recombination rate, $r_{nr}=\frac{\upsilon_sA}{V_a}n+Cn^3$, accounts for the surface recombination described by the surface velocity, $\upsilon_s$, and by the surface area of the active region, $A$, and for Auger recombination (described by $C$). The radiative decay into the leaky modes is calculated similarly as $r_{sp,cav}$ but replacing the cavity density of states by a density of leaky optical modes. The former is typically reduced from the bulk value in a nanophotonic cavity (see, e.g. in  micropillar cavities \cite{Gerard1998}) and can be calculated numerically \cite{Yamamoto1991}. The remaining parameters include $I$ describing the injection current, $q$ the electron charge, and $\eta_{i}$ the injection efficiency. Finally, the photon loss rate is given by $n_{ph}/\tau_p$, where $\tau_p=\frac{\lambda_c Q}{2\pi c}$ is the photon lifetime which is determined from the quality $Q-$factor and the cavity resonance wavelength, $\lambda_{c}$, where $c$ is the speed of light.

From the analysis of Eqs. (\ref{eq_rate_equations_1})-(\ref{eq_rate_equations_2}), we readily recognize that a substantial reduction of the mode volume results in an increase of the modal gain. This comes directly from the dependence of the stimulated emission rate on the field per photon. Importantly, this rate-equation model avoids the \emph{ad-hoc} introduction of the Purcell factor, $F_P$ , and the spontaneous emission factor, $\beta$, directly into the rate equations. This makes our theoretical treatment more transparent since only the physical parameters of the micro- or nano-cavity semiconductor laser such as the cavity dimensions and emitter/cavity relative linewidths are employed.

\subsection{Rate equations with spatially distributed emitter}

In the situation of a small-cavity laser with wavelength- or subwavelength-scale size, the spatial variation of the field in the active region of the small-cavity needs to be taken into account. In this case, the mode function $e(\overrightarrow{r})$, and correspondingly the transition rates, vary over the active volume. This case has been identified, for example, in the analysis of plasmonic nanolasers in \cite{Chang2009}, where an energy confinement factor was introduced in the rate equations to describe the non-perfect overlap of the optical mode with the gain medium. In our analysis, in order to account for this effect, we discretize the active volume in boxes, $dV_a$, small enough that the field $e(\overrightarrow{r})$ is approximately uniform within the box, and sufficiently large that a band description is adequate ($>40$ nm). Assuming that the carrier concentration is uniform over $V_a$ due to carrier diffusion, each box contributes to a cavity emission rate $\left(\gamma_{net}n_{ph}+\frac{\gamma_{sp,cav}}{V}\right)|e(\overrightarrow{r}_a)|^{2}dV_a$, and the integration over $V_a$ provides the following corrected expression for the effective mode volume:

\begin{equation}\label{eq_eff_volume}
  \frac{1}{V_{eff}}=\frac{1}{V}\int\frac{|e(\overrightarrow{r}_a)|^{2}}{V_a}dV_a
\end{equation}

From this definition, we recognize that, if the emitter-cavity overlap is not perfect, i.e., if the field amplitude is low in parts of the active region, the effective cavity volume is increased, and the spontaneous emission rate and gain are correspondingly decreased. This situation has been considered previously for the case of the spontaneous emission modelling in a metal-cavity nanoLED \cite{Dolores-Calzadilla2017} and also in a metallo-dielectric nanolaser \cite{Hill2007}, but it is here generalized to include also stimulated emission.

Using the corrected expression for the effective mode volume we arrive to the final rate-equation model for a nano-cavity laser:

\begin{eqnarray}
  \frac{dn}{dt} &=& \frac{\eta_{i} I}{q V_a}-r_{nr}-r_{l}-\frac{\gamma_{sp,cav}}{V_{eff}}-\gamma_{net}n_{ph} \label{eq_rate_equations_1_final}\\
  \frac{dn_{ph}}{dt} &=& \frac{V_a}{V_{eff}} \gamma_{net}n_{ph}+\frac{V_a}{V_{eff}} \frac{\gamma_{sp,cav}}{V_{eff}}-\frac{n_{ph}}{\tau_{p}}\label{eq_rate_equations_2_final}
\end{eqnarray}
where all the volume dependences are written explicitly, and $n_{ph}$ is now defined as $N_{ph}/V_{eff}$.

\section{Static properties of metallo-dielectric nanolasers}\label{sec_static}

Metallo-dielectric cavities can confine light to volumes with dimensions smaller than the wavelength. They typically consist of a semiconductor pillar (e.g. a double heterostructure InP/InGaAs/InP) surrounded by an isolating dielectric material (e.g. SiN) and then encapsulated with metal (gold or silver). The combination of metal and dielectric confines the optical mode around the semiconductor gain region. In this section, we consider three examples of metallo-dielectric micro- and nano-pillar cavity lasers and analyze their static properties using the rate-equation model presented in the previous section, Eqs. (\ref{eq_rate_equations_1_final})-(\ref{eq_rate_equations_2_final}). We note that the goal here is not to provide a comprehensive model including all relevant effects, but rather a simple and intuitive description of practical laser structures under realistic conditions. This will provide direct insight on the role of key parameters on the performance of the micro- and nano-lasers, particularly the surface recombination and mode volume.

\begin{figure}
  \centering
  \includegraphics[width=2.7in]{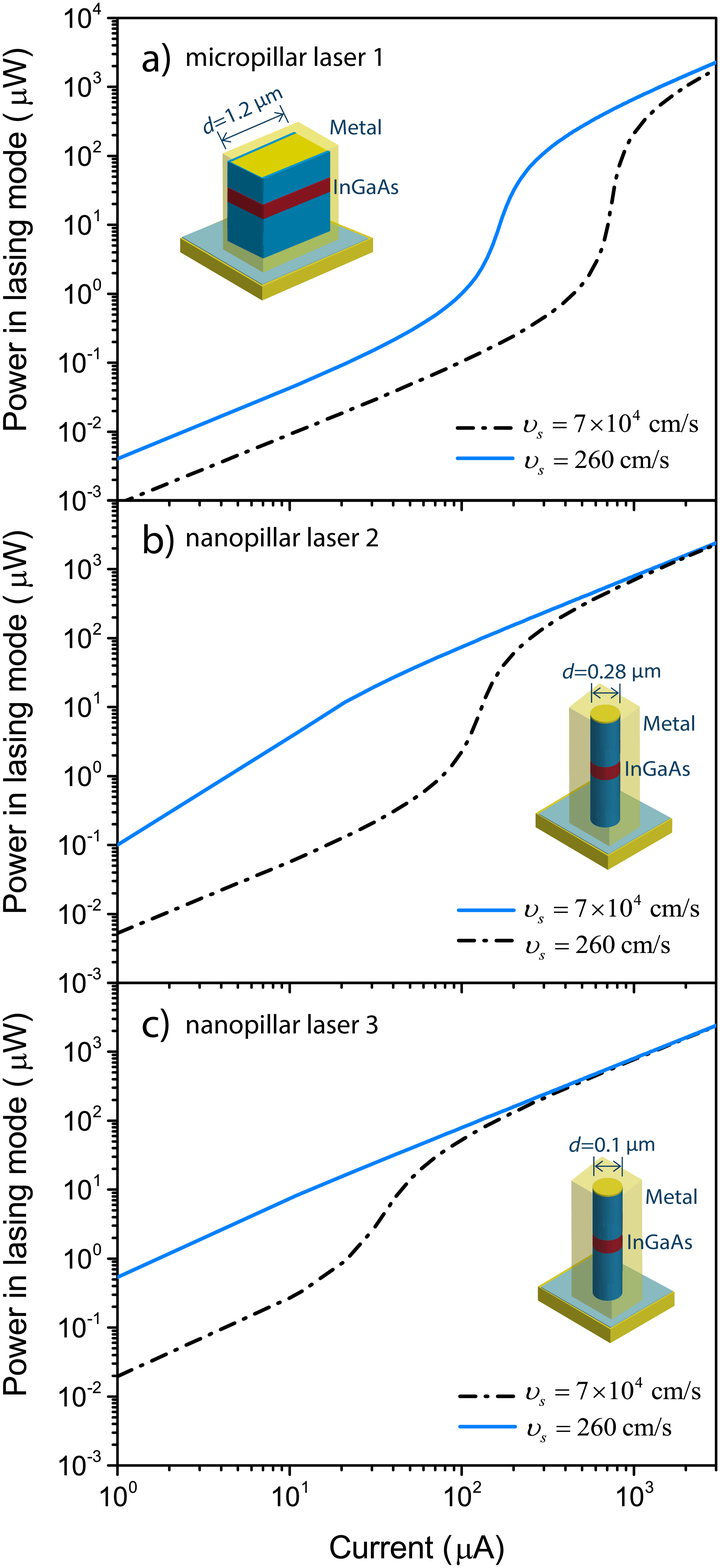}\\
  \caption{Simulated pillar $L-I$ characteristics of the metallo-dielectric micro- and nano-cavity lasers: a) micropillar laser 1 with a rectangular cross-section cavity pillar and a mode volume of $V_{eff}=0.5$ $\mu$m$^3$ (schematic of the micropillar shown in the inset); b) nanopillar laser 2 with a circular cross-section cavity pillar and a mode volume of $V_{eff}=0.025$ $\mu$m$^3$ (schematic of the nanopillar shown in the inset). c) nanopillar laser 3 with a circular cross-section cavity pillar and a mode volume of $V_{eff}=0.0025$ $\mu$m$^3$ (schematic of the nanopillar shown in the inset). The $L-I$ curves were simulated for the following values of surface recombination: $\upsilon_{s}=7 \times 10^4$ cm/s (dash-dot black trace) and $\upsilon_{s}=260$ cm/s (solid blue trace).}\label{fig_2}
\end{figure}

In the first metallic-coated micropillar laser analyzed here (shown in the inset of Fig. \ref{fig_2}a)), we assumed a pillar with rectangular cross-section (including the SiN dielectric layer) with dimensions of $1.15$ (width) $\times$ $1.39$ (length) $\times$ $1.7$ (height) $\mu$m$^3$. This structure is similar to the one presented in the work of Ding \emph{et al.} \cite{Ding13}, the first CW electrically-pumped nanolasers operating at room-temperature. The second metallic-coated nanopillar laser, corresponds to a pillar with a circular cross-section (shown in the inset of Fig. \ref{fig_2}b)). We assumed dimensions of $0.28$ $\mu$m (diameter) and $1.29$ $\mu$m (height). This structure is similar to the one presented in the work of Hill \emph{et al.} \cite{Hill2007}, the first reported metallic coated nanolaser. Lastly, in the third nanopillar laser analyzed here, we assumed a similar geometry as in the previous nanolaser with a diameter of $0.1$ $\mu$m (shown in the inset of Fig. \ref{fig_2}c)). In all three lasers, we assumed values of active volume, $V_a$, and effective mode volume, $V_{eff}$, such that $V_{eff}\sim V_a$, see Table I. We note that for the micropillar laser 1 and nanopillar laser 2, the effective mode volume is larger than the wavelength of the emission, whereas for the nanopillar laser 3 the mode volume is smaller than the wavelength. This nanolaser 3 corresponds to a Purcell enhanced case via the reduction of the effective mode volume of the nanocavity. Since in this situation the cavity is still larger than the wavelength at least in one direction, calculations of the relative fraction of magnetic energy show that more energy is stored in the magnetic field than in the motion of electrons in the metal \cite{Khurgin2014}, and therefore the mode volume definition of Eq. (\ref{eq_mode_volume}) and the effective mode volume definition of Eq. (\ref{eq_eff_volume}) are still approximately valid. In the case of metallo-dielectric cavities that are sub-wavelength in all three directions, the kinetic energy can be included in the calculation of the effective mode volume, as discussed in \cite{Khurgin2014}.

  \begin{table}
\caption{Description of the typical parameters used in the model.}
\label{tab:1}       
%
%
\begin{tabular}{p{2.7cm}p{1.55cm}p{1.55cm}p{1.55cm}}

\textbf{Parameter} & \textbf{micropillar laser 1} & \textbf{nanopillar laser 2} & \textbf{nanopillar laser 3} \\

Temperature, $T$ &  $300$ K  & $300$ K & $300$ K \\
Electron mass, $m_e$ &  $0.041$ $m_0$  & $0.041$ $m_0$ & $0.041$ $m_0$ \\
Hole mass, $m_h$ &  $0.45$ $m_0$  & $0.45$ $m_0$ & $0.45$ $m_0$ \\
Refractive index, $n_{ra}$ &  $3.4$ & $3.4$ & $3.4$ \\
Quantum efficiency, $\eta_i$ & $1.0$  & $1.0$ & $1.0$ \\
Active volume, $V_{a}$ & 0.4 $\mu$m$^3$  & 0.02 $\mu$m$^3$ & 0.002 $\mu$m$^3$ \\
Wavelength, $\lambda_{c}$ & 1.55 $\mu$m & 1.55 $\mu$m & 1.55 $\mu$m \\
Quality factor, $Q$ & 235  & 235 & 235 \\
Mode volume, $V_{eff}$ & 0.5 $\mu$m$^3$  & 0.025 $\mu$m$^3$ & 0.0025 $\mu$m$^3$ \\
Surface velocity, $\upsilon_{s}$ & $7 \times 10^4$ cm/s  & $7 \times 10^4$ cm/s & $7 \times 10^4$ cm/s  \\
Auger recombination, $C$ & $8.5 \times 10^{-29}$ cm$^6$/s  & $8.5 \times 10^{-29}$ cm$^6$/s & $8.5 \times 10^{-29}$ cm$^6$/s \\
\end{tabular}
\end{table}

We have numerically simulated the $L-I$ characteristics of the small lasers, Fig. \ref{fig_2}, using the rate-equation model described by Eqs. (\ref{eq_rate_equations_1_final})-(\ref{eq_rate_equations_2_final}). In all cases, the gain active medium was a bulk InGaAs material and we assumed room-temperature operation at $1.55$ $\mu$m. In the simulations, $\gamma_{net}$ and $\gamma_{rsp,cav}$ were numerically calculated employing typical values found in the literature for the InGaAs active material. In order to allow a direct comparison, we assumed a quality factor of $Q=235$ for all cavities, corresponding to a photon lifetime of $\tau_p=0.19$ ps. In practical structures, it is expected that $Q$ values $>200$ can be achieved at room-temperature using optimized metal layers \cite{Ding2013a}. Table \ref{tab:1} summarizes the parameter values used in the $L-I$ simulations. In order to plot the $L-I$ curves, we found the steady-state solutions of Eqs. (\ref{eq_rate_equations_1_final})-(\ref{eq_rate_equations_2_final}) by setting $dn/dt$ and $dn_{ph}/dt$ to zero and then solving the equations in the unknown $E_{f,c}$ and $E_{f,v}$. The Fermi distribution functions $f_c$ and $f_v$ were computed from the electron density, $n_{c}$ and hole density, $n_{h}$, related with the respective quasi-Fermi levels $E_{f,c}$ and $E_{f,v}$ and assuming the charge neutrality condition $(n_{c}=n_{h})$. The calculated photon density was then converted to an output power using $P=\frac{n_{ph}V_{eff}\eta hc}{\tau_p \lambda_{c}}$, where $h$ is the Planck's constant, and $\eta$ the external quantum efficiency. Lastly, for simplicity of analysis, the injection efficiency ($\eta_i=1$) and the external quantum efficiency ($\eta=1$) were kept constant.

\begin{figure}
  \centering
  \includegraphics[width=3.5in]{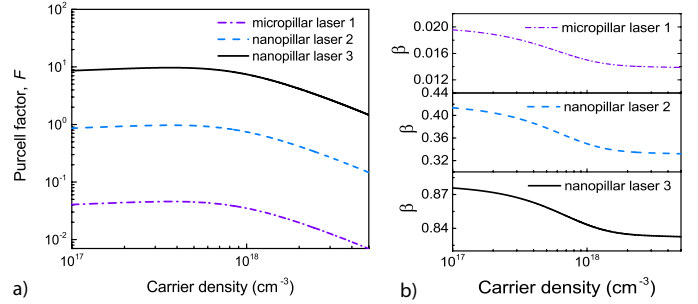}\\
  \caption{a) Purcell factor as a function of the carrier density for the metallo-dielectric micro- and nanocavity lasers shown in Fig. \ref{fig_2}. b) Corresponding theoretical values of $\beta$-factor as a function of the carrier density.}\label{fig_3}
\end{figure}

In Fig. \ref{fig_2} the calculated $L-I$ curves are displayed showing the optical power versus the injected current. The curves were simulated for the following values of surface recombination: $\upsilon_{s}=7 \times 10^4$ cm/s (dash-dot black trace), a typical value found in micro- and nanopillar devices \cite{Ding2013a,Dolores-Calzadilla2017}, and $\upsilon_{s}=260$ cm/s (solid blue trace), an ultralow value of surface recombination achieved recently in InGaAs/InP nanopillars using an improved passivation method \cite{Aura2017}. In all plots, we kept a realistic room-temperature Auger coefficient (see Table \ref{tab:1}). This choice of parameters results in a threshold current ($I_{th}=0.74$ mA) close to the value experimentally reported ($I_{th}\sim1$ mA) in a similar structure at room temperature \cite{Ding13}. The results in Fig. \ref{fig_2} show a substantial reduction of the laser threshold in the case of a low surface velocity for all small lasers, demonstrating that nonradiative effects play a strong role in the performance of the nanolasers. Furthermore, for the smallest cavities, panel b) and c), we also see a smooth transition from non-lasing to lasing for the case of low surface velocity recombination (blue solid curves). This effect is a result of the substantial reduction of the surface recombination together with the small mode volume where a substantial fraction of the spontaneous emission is coupled to the cavity mode below threshold. In the sub-wavelength case, Fig. \ref{fig_2}c), the threshold transition disappears completely in the $L-I$ curve. This is a case of a nanolaser exhibiting a 'threshold-less' behavior. As discussed next, the corresponding calculated values of $\beta$, Fig. \ref{fig_3}b), indeed show that the theoretical $\beta$ approaches unity when $V_{eff}$ is substantially reduced (assuming a fixed radiative emission into leaky modes, $r_l$).

It is noteworthy that the curves in Fig. \ref{fig_2} were simulated without requiring the introduction of the spontaneous emission factor, $\beta$, or the Purcell factor, $F_P$. In our model, the corresponding theoretical values of $\beta$ and Purcell enhancement $F$ can be calculated from Eqs. (\ref{eq_beta}) and (\ref{eq_purcell_bulk}), respectively, for a given active gain material and effective mode volume. Figure \ref{fig_3}a) shows the theoretical Purcell factor as a function of carrier density for all lasers displayed in Fig. \ref{fig_2}. Clearly, a decrease of the mode volume produces a proportional increase of the Purcell factor. Since we choose identical homogeneous and inhomogeneous broadening conditions and identical wavelength operation, the Purcell factor scales as $1/V_{eff}$. The calculations predict a Purcell factor of $F\sim4.2$ for the case of the smallest nanopillar laser 3, and no enhancement, that is $F<1$, for the remaining cases (all values taken at a carrier density of $2 \times 10^{18}$ cm$^{-3}$). The Purcell factor strongly depends on the carrier density. The large decrease of Purcell enhancement for carrier densities above the transparency value is a combination of band filling effect and the continuous increase of $R_{bulk}$.

The values of the Purcell factor calculated here, $F<10$, for the smallest mode volume analyzed and Purcell factors below one, for the larger mode volumes, are substantially lower than the values typically employed in the numerical fittings reported elsewhere using rate-equation analysis of experimental nanolasers with similar cavity dimensions and bulk gain medium (see e.g. \cite{Ding13}). While indeed calculations of Purcell enhancement (e.g. using finite-difference time-domain simulations) can show very high Purcell values ($F>10$) for the mode of interest in the ideal case of a monochromatic dipole, our theoretical model shows that this value is substantial reduced in the bulk case when homogeneous and inhomogeneous broadening are taken into account. Indeed, the strong reduction of the Purcell enhancement due to the broadening effects has been recognized in a previous theoretical work that analyzed a nanolaser with MQW active gain medium \cite{Suhr2010}. Importantly, our model also shows that the non-perfect spatial overlap of the optical mode with the gain medium further contributes to the reduction of the maximum achievable Purcell enhancement.

Finally, the $\beta$ values are shown in Fig. \ref{fig_3}b) as a function of the carrier density calculated using Eq. (\ref{eq_beta}) (we assumed an emission rate into the leaky modes fixed to the value used in the $L-I$s shown in Fig. \ref{fig_2}). As the mode volume decreases, the $\beta-$factor quickly approaches unity, that is, a substantially large portion of the spontaneous emission is emitted in the lasing mode. The varying $\beta$ can be an important feature in cases where detailed studies of the nanolaser properties below and around threshold are required. This is significantly different from the standard rate-equation analysis where $\beta$ is assumed constant.

\section{Dynamic properties of metallo-dielectric nanolasers}\label{sec_dynamic}

While a wide range number of metallo-dielectric nanolasers, similar to the ones described in the previous section, have been reported, their respective modulation properties remain largely unknown, namely because of the typical ultralow output power levels. Here, we perform a systematic study to investigate the influence of the substantial reduction of the mode volume in the expected performance of the nanolasers in terms of modulation speed.

\begin{figure}
  \centering
  \includegraphics[width=2.5in]{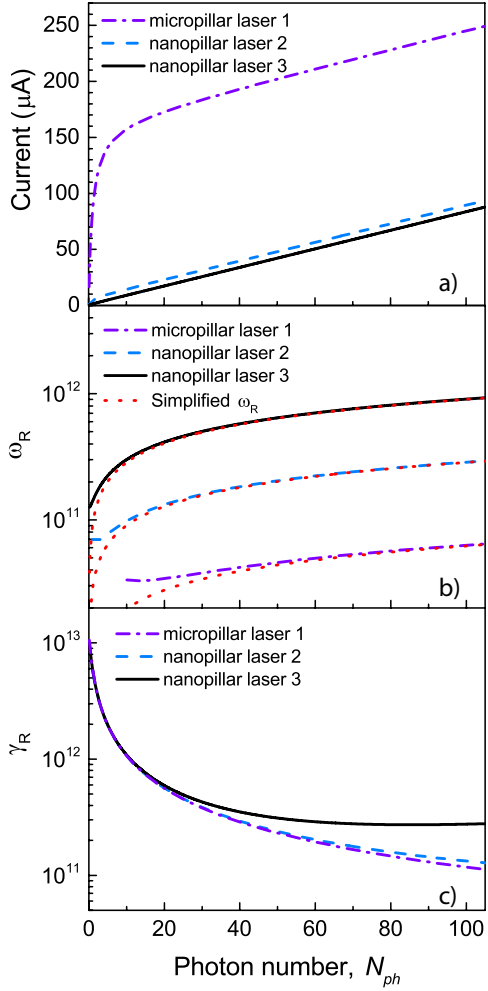}\\
  \caption{a) The injected current, b) the relaxation oscillation frequency, $\omega_R$, and c) and the damping, $\gamma_R$ as a function the photon number for the metallo-dielectric micro- nano-cavity lasers shown in Fig. \ref{fig_2}. In panel b), the dot red curves correspond to the simplified expression for the relaxation oscillation frequency, $\omega_R\approx \sqrt{\frac{N_{ph}}{\tau_p V_{eff}} \frac{\partial\gamma_{net}}{\partial n}}$.}\label{fig_4}
\end{figure}

Figure \ref{fig_4}a) displays the injected current versus the photon number for the metallo-dielectric nanolasers analyzed in Fig. \ref{fig_2}, in the case of low surface recombination velocity. From the $I-L$ curves, we immediately see that the metallo-dielectric cavity nanolasers operate with much lower number of photons than standard semiconductor lasers (typically two order of magnitude lower than a typical small VCSEL), explaining the ultralow output power levels usually reported and the difficulties in measuring the respective modulation properties. Although further increase of the current would allow us to increase the photon number output, in realistic devices, effects such as the temperature increase (not analyzed here)\cite{Smalley2014}, and Auger recombination strongly limit the current range in which these devices can be operated.

Using a standard small-signal analysis of the differential equations, Eqs. (\ref{eq_rate_equations_1_final})-(\ref{eq_rate_equations_2_final}) (see Appendix A for more details), we analyzed the modulation characteristics of the devices of Table I and Fig. \ref{fig_2}, specifically the relaxation oscillation frequency, Fig. \ref{fig_4}b), and the damping factor, Fig. \ref{fig_4}c), as a function of the photon number. As discussed in the Appendix A, the relaxation oscillation frequency, $\omega_R$, of the nanolasers is given approximately by $\omega_R\approx \sqrt{\frac{N_{ph}}{\tau_p V_{eff}} \frac{\partial\gamma_{net}}{\partial n}}$, see dot red curves in Fig. \ref{fig_4}b), which agrees with the $\omega_R$ expression found in laser textbooks \cite{Coldren2012}. While for lower photon numbers the spontaneous emission into the cavity also contributes to $\omega_R$ (Eq. (\ref{eq_relaxation_oscillation_appr}) in the Appendix A), in Fig. \ref{fig_4}b) we see that at values of photon number $N_{ph}>20$, the simplified expression is sufficient to describe the relaxation oscillation predicted by the full model. Our results clearly demonstrate that the modulation dynamics for bias values well above the threshold depends on $V_{eff}$ through the gain term as typically observed in a standard laser and is not affected by the spontaneous emission term. This explains the higher slope of $\omega_R$ for decreasing effective mode volume of the nanolasers shown in Fig. \ref{fig_4}.

\begin{figure}
  \centering
  \includegraphics[width=3.6in]{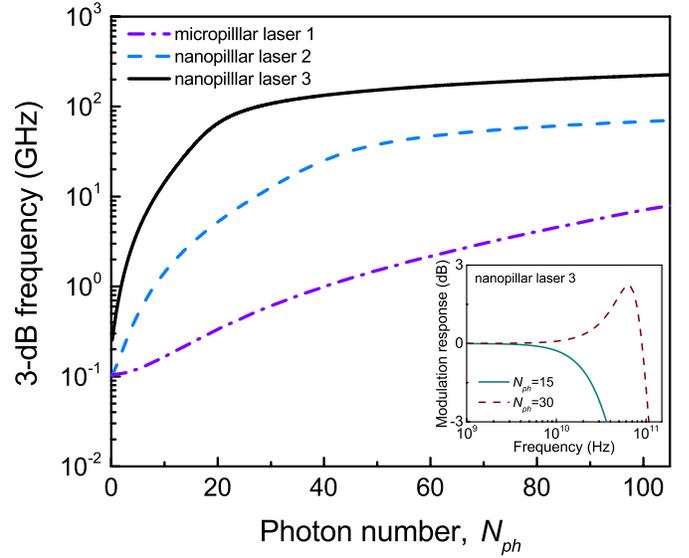}\\
  \caption{The small-signal 3dB-bandwidth ($f_{3dB}$) versus photon number for the metallo-dielectric micro- nano-cavity lasers analyzed in Fig. \ref{fig_4}. In the inset is shown the modulation response of nanopillar laser 3 as a function of the frequency at $N_{ph}=15$ and $N_{ph}=30$.}\label{fig_5}
\end{figure}

In the case of the damping factor, Fig. \ref{fig_4}c), we note that the plots are very similar (note that the $y$-axis is in log scale) since we have chosen cavities with the same quality factor. We also note a large value of the damping due to the low$-Q$ of the cavities, and a large increase of the damping for low photon number, i.e., $N_{ph}<20$. This pronounced variation of the damping close to threshold can be explained by a smooth increase of carrier density in the photon range between $N_{ph}=1$ and $N_{ph}=20$. This occurs since near and above the threshold region the damping factor decreases as $\frac{1}{\tau_p} - \gamma_{net}(n_0)$ (see Eq. (\ref{eq_damping_appr}) in Appendix A), where $n_0$ in $\gamma_{net}(n_0)$ is the carrier density value at the steady-state. Since in this region the carrier density is not fully clamped and the net gain increases smoothly, the large variation of the damping is more pronounced in the case of nanolasers, although it is also expected in micro- and macro-scale lasers when $N_{ph}$ is small. For $N_{ph}$ between 20 and 100 we note that the damping of the nanopillar laser 3 is slightly larger than the remaining lasers. This is a direct consequence of the contribution of $\tau_p\omega_{R}^{2}$ (see Eq. \ref{eq_damping} in the Appendix A) to the damping due to the large value of the relaxation oscillation frequency ($\omega_{R} \sim 9 \times$ 10$^{11}$ at $N_{ph}=100$) in the case of the nanopillar laser 3. Lastly, we note that the increase of $\gamma_{R}$ with $N_{ph}$, which is typical of larger lasers \cite{Coldren2012}, will not be likely observed in the nanolasers analyzed here due to the low achievable photon numbers. This fact, together with the low $Q-$factor and the incomplete carrier clamping at threshold, makes the current dependence of the damping factor in nanolasers markedly different from the one in larger lasers.

In Fig. \ref{fig_5} we show the calculated small-signal 3dB-bandwidth as a function of the photon number (see Appendix A). The 3dB-bandwidth plot clearly shows a large increase of the modulation speed well above 100 GHz for the case of the smallest nanopillar laser 3, as compared with the micropillar laser 1  showing a modulation bandwidth close to 10 GHz for $N_{ph}=100$. This allows us to conclude that a large increase of speed in nanolasers can be achieved as a direct consequence of the strong reduction of the effective mode volume and corresponding enhancement of the stimulated emission rate. The modulation response for nanopillar laser 3 is shown in the inset of Fig. \ref{fig_5} and allow us to explain the different slopes observed in the 3-dB frequency curves for nanopillar lasers 2 and 3. This change of slope marks the transition of the nanolaser between an overdamped regime typical of a low-$Q$ laser oscillator, that is, when the relaxation oscillation signature is absent (in the inset for $N_{ph}=15$), and an underdamped regime with a clear relaxation oscillation frequency signature characteristic of standard lasers (in the inset for $N_{ph}=30$). Effects such as temperature increase strongly limits the current range in which a practical nanolaser can be operated. Therefore, it is expected that realistic nanolasers will operate mostly in the overdamped regime since a large current density ($>200$ kA/cm$^{2}$) would be required in order to operate the nanolasers in the standard underdamped regime.

\section{Conclusion}\label{sec_conclusion}

In this work, we have investigated the role of Purcell effect in the stimulated and spontaneous emission rates of semiconductor lasers over a wide range of cavity dimensions and emitter/cavity relative linewidths using single-mode rate-equation analysis. We extended our treatment to account for the inhomogeneous broadening of the carriers and their spatial distribution over the volume of the active region enabling the detailed modeling of either micro- or nano-scale lasers. Using this model, we have investigated the static and dynamic characteristics of wavelength- and sub-wavelength scale electrically-pumped metallo-dielectric cavity nanolasers. The ultimate limits of scaling down these nanoscale light sources leading to Purcell enhancement of the emission and higher modulation speeds were discussed. We have shown that the modulation dynamics depend directly on the effective mode volume, $V_{eff}$, and the photon number, $N_{ph}$, through the gain terms and is not significantly affected by the spontaneous emission terms. As a result, the ultrafast modulation speed properties predicted in nanolasers are a direct consequence of the enhancement of the stimulated emission rate via reduction of the mode volume.

The treatment presented here is markedly distinct from the rate-equation analysis reported elsewhere due to the combination of the following key characteristics: i) only the physical properties of the nanolasers, specifically the gain material and cavity characteristics, are sufficient to fully describe their static and dynamic characteristics; ii) both spontaneous and stimulated emission rates are equally treated which leads to a Purcell enhancement of both radiative emissions; and  iii) the equations do not require the \emph{ad-hoc} introduction of the spontaneous emission factor, $\beta$, or the Purcell factor, $F_P$, frequently adopted as fitting parameters in the rate-equation models. Furthermore, if required our model can be extended to include additional details, namely a more detailed description of the semiconductor band structure, the description of gain compression effects, or the inclusion of thermal effects \cite{Smalley2014}, which can be highly important for the future design of nanolasers and respective prediction of their performance. Specifically, peculiar effects such as a non-monotonic dependence on temperature of the spontaneous emission factor as reported in \cite{Smalley2014} and their respective impact in the dynamic properties of nanolasers would be interesting to study using our model. The theoretical analysis presented here is important for the study of nanoscale semiconductor light sources and their realistic performance for applications in future nanophotonic integrated circuits.

\appendices
\section{Modulation response}

To obtain the high-speed modulation response, we perform a small-signal analysis following a standard procedure \cite{Coldren2012} by taking the total differential of the rate equations Eqs. (\ref{eq_rate_equations_1_final})-(\ref{eq_rate_equations_2_final}):

\begin{eqnarray}
  \frac{d}{dt}(dn) &=& \frac{\eta_{i} dI}{q V_a}- \gamma_{nn}dn - \gamma_{np}dn_{ph} \label{eq_rate_equations_1_differential}\\
  \frac{d}{dt}(dn_{ph}) &=& \gamma_{pn}dn - \gamma_{pp}dn_{ph} \label{eq_rate_equations_2_differential}
\end{eqnarray}

Where the coefficients can be written as:

\begin{eqnarray}
  \gamma_{nn} &=& \frac{\partial r_{nr}}{\partial n} + \frac{\partial r_{l}}{\partial n} + \frac{1}{V_{eff}}\frac{\partial\gamma_{sp,cav}}{\partial n} + n_{ph}\frac{\partial \gamma_{net}}{\partial n}\label{eq_gamma_nn1}\\
  \gamma_{np} &=& \gamma_{net}(n_0) \label{eq_gamma_np1}\\
  \gamma_{pn} &=& \frac{V_a}{V_{eff}^{2}}\frac{\partial\gamma_{sp,cav}}{\partial n}+\frac{V_a}{V_{eff}}n_{ph}\frac{\partial \gamma_{net}}{\partial n}  \label{eq_gamma_pn1}\\
  \gamma_{pp} &=& \frac{1}{\tau_{p}}-\frac{V_a}{V_{eff}}\gamma_{net}(n_0)  \label{eq_gamma_pp1}
\end{eqnarray}
where $n_0$ in $\gamma_{net}(n_0)$ is the carrier density value at the steady-state. We note that in the analysis used here the intraband dynamics and thereby gain compression is neglected (i.e. $\gamma_{net}$ does not depend on $n_{ph}$).

To obtain the small-signal responses $dn(t)$ and $dn_{ph}(t)$ to a sinusoidal current modulation $dI(t)$, we assume solutions of the form $dI(t)=\Delta Ie^{i\omega t}$, $dn(t)=\Delta n e^{i\omega t}$ and $dn_{ph}=\Delta n_{ph}e^{i\omega t}$. Following the standard procedure we apply Cramer’s rule to obtain the small-signal carrier and photon densities in terms of the modulation current. The modulation transfer function is then given by:

\begin{equation}\label{eq_modulation_transfer_function}
  H(\omega)=\frac{\omega^2_R}{\omega^2_R-\omega^2+i\omega\gamma_{R}}
\end{equation}
where $\omega_R$ is the relaxation resonance frequency and $\gamma_{R}$ the damping factor and are they are related to the coefficients as:

\begin{eqnarray}
  \omega^2_R &=& \gamma_{nn}\gamma_{pp}+\gamma_{pn}\gamma_{np}\label{eq_relaxation_oscillation}\\
  \gamma_{R} &=& \gamma_{nn}+\gamma_{pp}\label{eq_damping}
\end{eqnarray}

In the situation where a) the nonradiative contribution can be neglected, b) $V_{eff}\sim V_a$, as in the case of subwavelength nanolasers, and c) the contribution of the leaky modes can be neglected (e.g. in a high-$\beta$ nanolaser where the radiative emission into the lasing mode is large), $\omega^2_R$ and $\gamma_{R}$ can be approximated as:

\begin{eqnarray}
  \omega^2_R &\simeq& \frac{N_{ph}}{\tau_p V_{eff}}\left( \frac{1}{N_{ph}}\frac{\partial\gamma_{sp,cav}}{\partial n}+\frac{\partial\gamma_{net}}{\partial n} \right) \label{eq_relaxation_oscillation_appr}\\
  \gamma_{R} &\simeq& \tau_p \omega_R^{2} + \frac{1}{\tau_p} - \gamma_{net}(n_0) \label{eq_damping_appr}
\end{eqnarray}
Whereas for a low photon number Eq. (\ref{eq_relaxation_oscillation_appr}) shows a dependence on both differential spontaneous emission and net gain, when the photon number is large Eq. (\ref{eq_relaxation_oscillation_appr}) simplifies to $\omega^2_R\approx \frac{N_{ph}}{\tau_p V_{eff}} \frac{\partial\gamma_{net}}{\partial n}$, which agrees with the typical expression found in laser textbooks \cite{Coldren2012}. This dependence of $\omega^2_R$ on the inverse of $V_{eff}$ clearly demonstrates that the modulation dynamics depends on the photon lifetime and on $V_{eff}$ and $N_{ph}$ through the gain term and is not affected by the spontaneous emission term. In the case of the damping factor, two regimes are distinguished: a) near and above the threshold region where the photon number is very low ($N_{ph}<10$ for the lasers analyzed here), the damping factor decreases as $\frac{1}{\tau_p} - \gamma_{net}(n_0)$; b) for larger photon number, the contribution of the term $\tau_p \omega_R^{2}$ in Eq. (\ref{eq_damping_appr}) becomes relevant. This can be seen in the plots of the Fig. \ref{fig_4}c) where the damping of the nanopillar laser 3 is slightly larger than the remaining lasers. The increase of $\gamma_{R}$ with $N_{ph}$ typical of larger lasers is not observed in the results shown in Fig. \ref{fig_4}c) due to the low achievable photon numbers.

Lastly, the 3-dB modulation bandwidth is given by:

\begin{equation}\label{eq_3dB_frequency}
  f_{3dB}=\frac{1}{2\pi}\sqrt{\omega^2_R-\frac{\gamma^2_R}{2}+\sqrt{\left(\omega^2_R-\frac{\gamma^2_R}{2}\right)^2+\omega^4_R}}
\end{equation}

\section*{Acknowledgment}


The authors would like to thank Meint Smit, Eindhoven University of Technology, for fruitful discussions on metallo-dielectric nanolasers, Jesper M{\o}rk, Technical University of Denmark, for useful discussions on the Purcell effect in nanolasers, and Julien Javaloyes, University of Balearic Islands, for discussions on numerical simulations and laser dynamics.

\ifCLASSOPTIONcaptionsoff
  \newpage
\fi

\end{document}